# Ultrahigh Thermal Conductivity in Isotopically Ultrapure Single-Crystal Diamond

Jinwen Liu[1], Junliang Fan[2], Zifeng Huang[2],Zhe Cheng[1,2,*]

1 School of Software & Microelectronics, Peking University, Beijing 100871, China

2 School of Integrated Circuits and Beijing Advanced Innovation Center for Integrated Circuits, Peking University, Beijing 100871, China

*Author to whom correspondence should be addressed: zhe.cheng@pku.edu.cn

## Abstract

Diamond has the highest thermal conductivity among natural materials. Reducing isotope scattering can further push the limit of thermal conductivity of diamond. Here, we report the room-temperature thermal conductivity of a diamond single crystal with an exceptionally low ppm-level $^{13}$C content. Secondary ion mass spectrometry (SIMS) reveals a $^{13}$C concentration of only ~0.00024% and negligible other impurities, placing this sample as the most isotopically purified diamond samples reported to date. Time-domain thermoreflectance measurements show a room-temperature thermal conductivity of ~3100 W $m^{-1}$ $K^{-1}$. Although this value is markedly higher than that of natural diamond, it remains comparable to the best reported values for less isotopically $^{12}$C pure diamond. This result suggests that additional phonon-scattering mechanisms beyond isotope scattering may exist at this extreme isotopic purification level. Vacancy defects may contribute to the remaining suppression of thermal conductivity.

## Introduction

Understanding the upper limit of heat conduction in three-dimensional solids is a central topic in thermal transport.[1–6] Owing to its low carbon mass, strong covalent bonding, and simple crystal structure, natural diamond exhibits one of the highest thermal conductivities among bulk materials (~2200 W $m^{-1}$ $K^{-1}$).[7] In semiconductors, heat is carried predominantly by phonons, and the thermal conductivity is determined by both the phonon spectrum and the relaxation time of phonon scatterings. The latter are limited by intrinsic Umklapp phonon-phonon scattering as well as scattering from isotope, impurities, grain boundaries, sample boundaries, and other defects.[8-10] Among these scattering mechanisms, carbon isotope scattering is particularly relevant to diamond. Natural diamond contains approximately 1.07% $^{13}C$,[11] and the mass difference between $^{12}C$ and $^{13}C$ gives rise to phonon-isotope scattering, thereby reducing the lattice thermal conductivity. Therefore, lowering the $^{13}C$ concentration is an effective route to suppress isotope scattering and enhance phonon thermal transport.[12]

Previous theoretical and experimental studies have proven that isotopically $^{12}C$ pure diamond can exhibit room-temperature thermal conductivity above 3000 W $m^{-1}$ $K^{-1}$, which is ~50% higher than that of natural diamond.[6,11–15] However, experimental data on diamond with ppm-level residual $^{13}C$ content remain limited, especially for samples that also contain ultralow concentration of other impurities (such as nitrogen). For such highly purified diamond, the actual thermal conductivity is still elusive.

In this work, we report the room-temperature thermal conductivity of a $^{12}C$ isotopically ultrapure single-crystal diamond with exceptionally low impurity concentration, measured by time-domain thermoreflectance (TDTR). Raman spectroscopy was used to confirm the characteristic diamond phonon mode, while secondary ion mass spectrometry (SIMS) was employed to verify the isotopic and impurity concentrations in the sample. By combining the measured thermal conductivity with the isotope composition and comparison with previously reported data, we assess the thermal transport characteristics of diamond in the extreme isotope-purity regime and discuss the possible origins of the remaining suppression of thermal conductivity.

**Results**

The $^{12}C$ isotopically purified single-crystal diamond sample investigated in this work was grown by Element Six and has a size of 4 mm × 4mm, with a total thickness of 600 μ m. To determine the thickness of the isotopically enriched region, we performed SIMS depth profiling and conducted Raman mapping across the polished sidewall for double check. As shown in Fig. 1a, the $^{13}C$ concentration remains at approximately the ppm level from the sample surface to a depth of ~17 μm, followed by a sharp increase. This indicates that the strongly $^{12}C$-enriched region extends approximately 17 μm from the surface.

As complementary evidence, Raman mapping provides a spatial identification of the boundary. As shown in Fig. 1b, the first-order diamond Raman peak remains at approximately 1332.25-1332.30 $cm^{-1}$ within the enriched region and shifts toward lower wavenumbers near the boundary with the substrate. Because Raman phonon frequency is sensitive to diamond's carbon isotope composition, the observed peak shift identifies the transition from the $^{12}C$-enriched layer to the underlying diamond substrate.[16,17] The Raman mapping gives an enriched-region thickness of approximately 20 μm, in reasonable agreement with the ~17 μm depth obtained from SIMS.

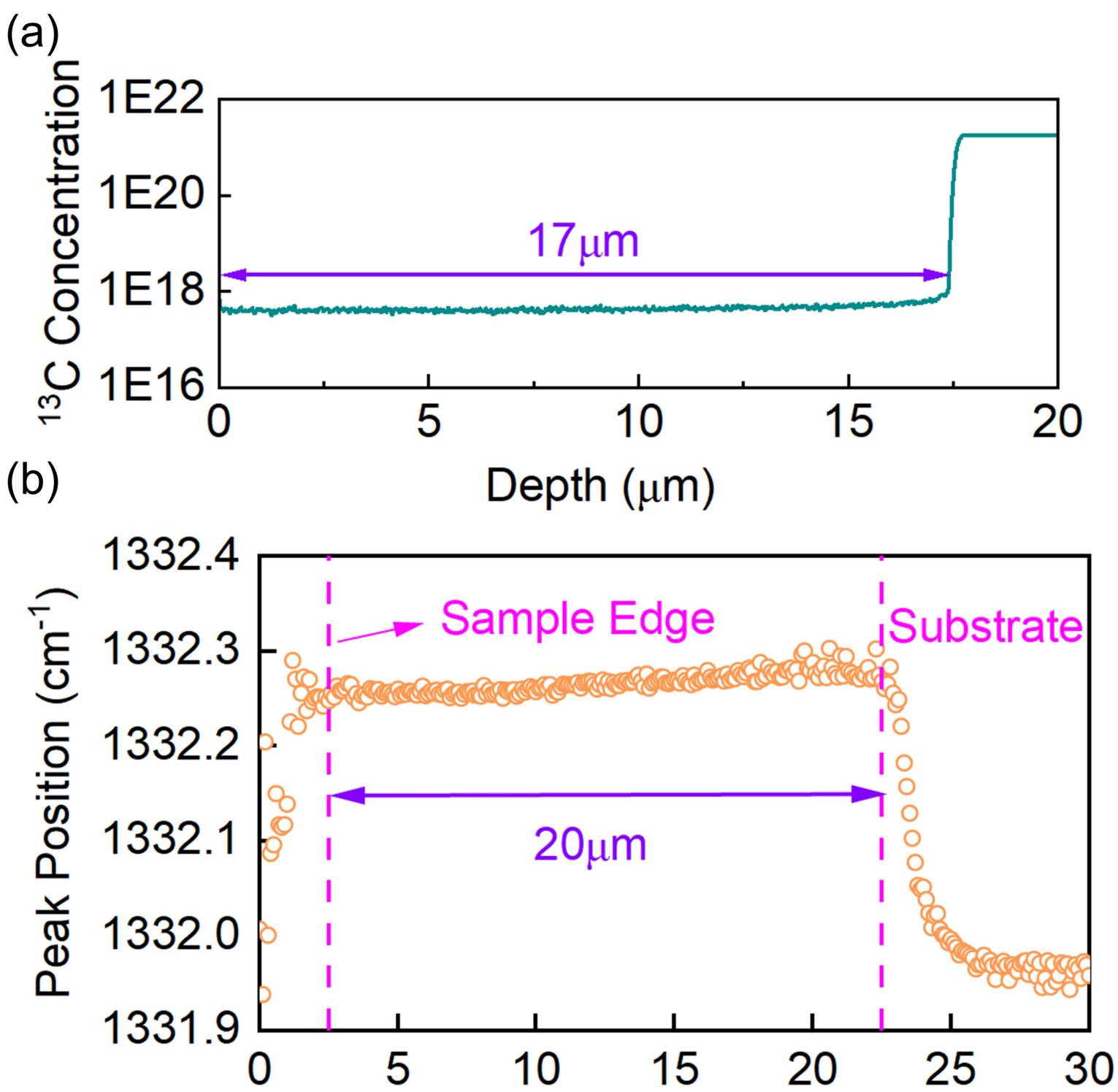

Figure 1. Characterization of the isotopically enriched region's thickness of the sample. (a) SIMS depth profile of 13C concentration showing the result increases sharply at a depth of ~17 μm, which indicates the strongly 12C-enriched region's thickness. (b)

Raman mapping across the polished sidewall. The Raman peak remains relatively stable within the enriched region and shifts toward lower wavenumbers at diamond substrate, corresponding to an enriched-layer thickness of approximately 20 μm.

Figure 2a shows the Raman spectrum of the sample measured using a 405 nm excitation laser and an 1800 grooves $mm^{-1}$ grating. A sharp first-order peak is observed at 1331.8 $cm^{-1}$, with a full width at half maximum (FWHM) of 3.0 $cm^{-1}$. The narrow Raman linewidth and the clear diamond phonon mode confirm the characteristic $sp^3$-bonded diamond lattice and indicate good crystalline quality of the layer.[18]

SIMS was used to determine the isotopic composition and quantify impurity content in the sample. As shown in Fig. 2b, the measured $^{13}C$ concentration is approximately ~0.00024%, which, to the best of our knowledge, is the lowest value reported to date. The SIMS depth profiles of H, O, Si, N, and $^{11}B$ are also shown in Fig. 2b. These signals are close to the detection limit over the measured depth range, indicating that the sample has a high chemical purity within the sensitivity of the SIMS measurement. These characterizations establish the exceptionally high isotopic and chemical purity of the diamond layer, providing a suitable precondition for probing thermal transport in the extreme isotope-purity regime.

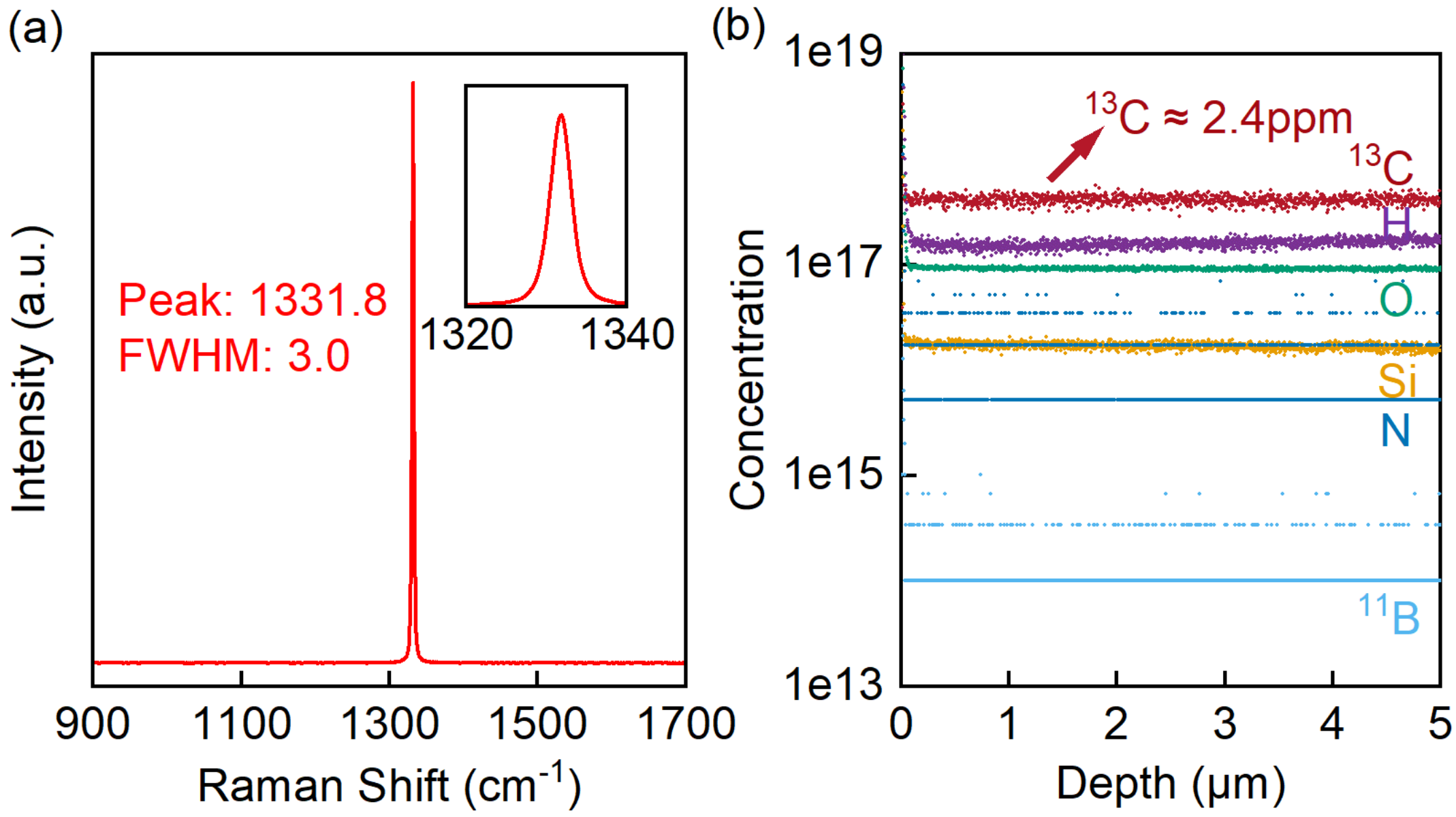

Figure 2. Structural characterization of the sample. (a) Raman spectrum of the sample showing a first-order diamond Raman peak at 1331.8 $cm^{-1}$ with a FWHM of 3.0 $cm^{-1}$. (b) SIMS depth profiles of $^{13}C$ and representative impurity elements in the sample. The ultralow $^{13}C$ concentration confirms the high $^{12}C$ purity of the sample.

The thermal conductivity (κ) of the sample was measured by TDTR. Prior to TDTR, an ~80 nm aluminum transducer layer was deposited on the sample surface by magnetron sputtering as transducer. The κ of the aluminum transducer was measured by a four-probe method along with the Wiedemann-Franz Law. Both two-color and two-tint setup TDTR configurations were used to check the consistency of the extracted thermal conductivity.[19,20]

The TDTR ratio signal, defined as $-V_{in}/V_{out}$, was analyzed using a multilayer heat diffusion model.[17] The thermal penetration depth is calculated as

$$d_p = \sqrt{\frac{\kappa}{\pi C f}} \quad (1)$$

where $C$ is the volumetric heat capacity, and $f$ is the modulation frequency. Using $\kappa$ = 3100 W m$^{-1}$ K$^{-1}$, $f$ = 5.02 MHz, and $C$ = 1.75*10$^{6}$ J m$^{-3}$ K$^{-1}$, the calculated $d_p$ is approximately 10.6 μm. This value is much smaller than the thickness of the $^{12}$C-enriched diamond layer, which is about 17 μm. Therefore, the TDTR signal is expected to be mainly sensitive to heat diffusion within the $^{12}$C-enriched layer rather than the underlying substrate (electronic grade single crystal diamond).

Figure 3a shows a representative TDTR ratio signal together with the best-fit curve. The good agreement between the experimental data and the model indicates that the heat diffusion process in the sample is well captured by the TDTR analysis. The dashed lines in Fig. 3a show the calculated ratio signals when the diamond thermal conductivity is varied by ± 10%, illustrating the sensitivity of the TDTR signal to the sample's thermal conductivity in the fitted delay-time range.[1]

The extracted room-temperature κ of the sample is ~3100 W m$^{-1}$ K$^{-1}$. TDTR measurements were performed at different modulation frequencies and laser spot sizes, as shown in Fig.3b. The fitted κ shows no clear dependence on modulation frequency. In contrast, a weak spot-size dependence is observed, with slightly lower apparent κ obtained at smaller laser spot sizes. This is due to the quasi-ballistic phonon transport effect, in which phonons with mean free paths longer than the spot size are not involved in the thermal transport.[21,22,23] The blue symbols in Fig.3b represent thermal boundary conductance (TBC) of Al-$^{12}$C ultrapure diamond interface, indicating the high quality of the sample surface and Al transducer.

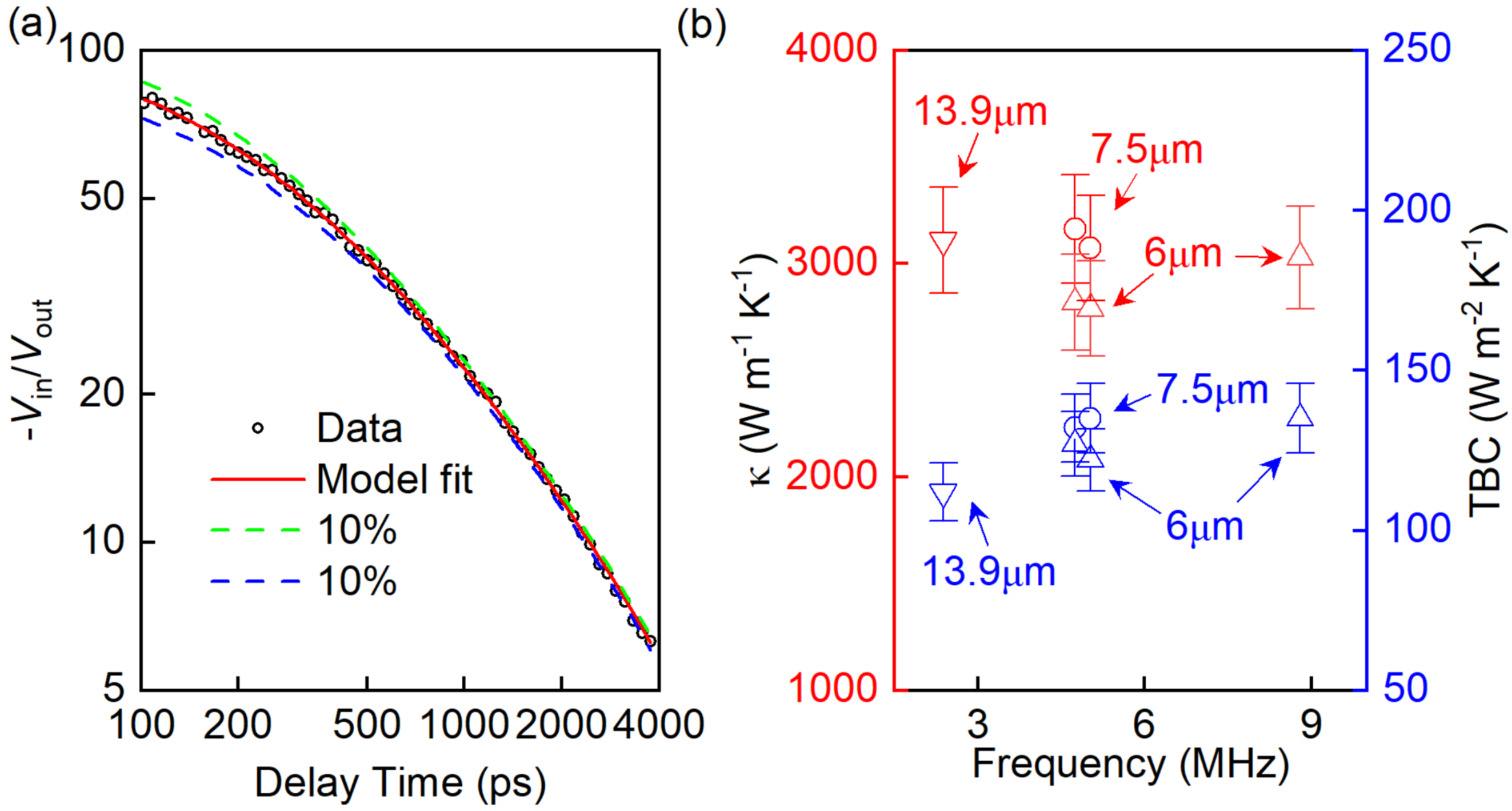

Figure 3. TDTR measurement of diamond sample. (a) Representative TDTR ratio signal and best-fit curve. The dashed curves were calculated with ± 10% changes in the diamond thermal conductivity, illustrating the sensitivity of the TDTR signal to the κ of the sample. (b) Extracted κ and TBC of Al-$^{12}$C ultrapure diamond interface as functions of modulation frequency under different laser spot-size conditions. The red symbols represent κ, while the blue ones represent TBC.

To place the present measurement in the context of isotope-dependent thermal transport, we compare the isotope composition and thermal conductivity of our sample with previously reported isotopically $^{12}$C pure diamond samples (Fig. 4).[11,13,16,24,25] The residual $^{13}$C concentration of our sample is only ~0.00024%, which is among the lowest residual $^{13}$C levels for diamond samples studied for thermal transport. The measured room-temperature κ of ~3100 W $m^{-1}$ $K^{-1}$ is significantly higher than typical values of natural diamond, confirming the high heat conduction capability of the sample.

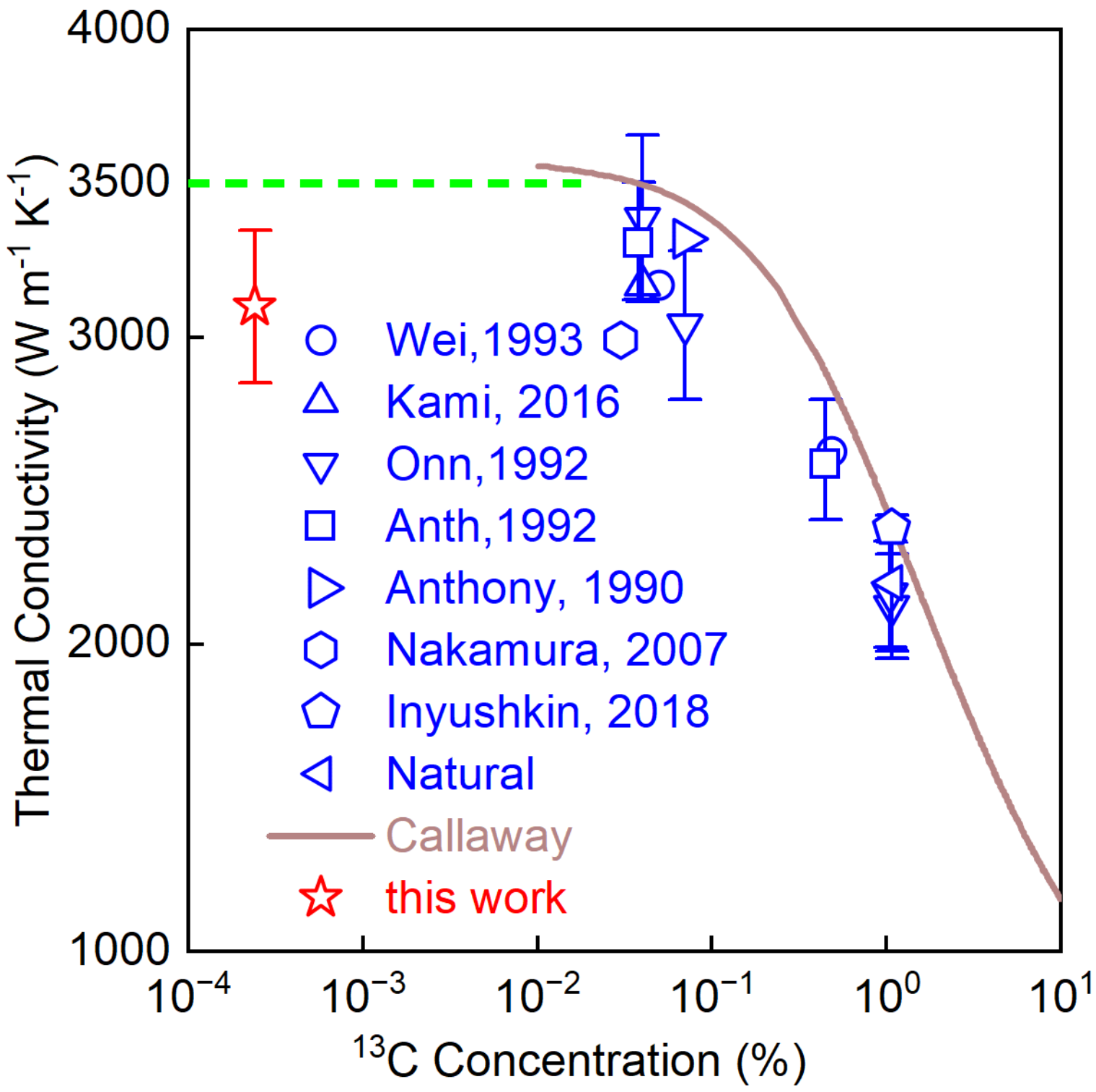

Figure 4. Comparison of room-temperature thermal conductivity of our sample as a function of $^{13}C$ concentration. Literature data for $^{12}C$ pure diamond samples are included for comparison in blue symbols[11,13,14,16,24-26] and green line[27]. The present sample, marked by the red star, has a $^{13}C$ concentration of ~0.00024% and a measured room temperature κ of ~3100 W $m^{-1}$ $K^{-1}$.

However, despite the extremely low $^{13}C$ content, the measured κ remains below the ~3500 W $m^{-1}$ $K^{-1}$ level expected for ideal $^{12}C$ diamond based on Callaway model and first principle calculations.[6,27] The comparison suggests that the enhancement of thermal conductivity with increasing isotopic purity does not always maintain remarkable. As $^{13}C$ scattering is progressively suppressed, other phonon-scattering processes may increasingly influence the measured thermal conductivity. Based on previous SIMS and Raman results, high crystalline quality is confirmed and prominent isotopic and impurity scattering are excluded. Thus, a plausible source is vacancy defects. Ab initio Green's function calculations have shown that vacancies in diamond can strongly scatter phonons because they substantially perturb the local harmonic interatomic force constants (IFCs),[28] while electron-irradiation experiments have directly linked vacancy concentration to the reduction of κ of diamond.[29] Therefore, residual vacancy defects, which are not directly quantified by SIMS, may contribute to the remaining suppression of κ in the present isotope ultrapure diamond.

**Summary**

In conclusion, we measured the thermal conductivity of an isotopically ultrapure $^{12}C$ diamond using TDTR. The sample has a $^{13}C$ concentration of ~0.00024% and a room temperature thermal conductivity of ~3100 W $m^{-1}$ $K^{-1}$. The κ shows no clear dependence on modulation frequency, while its weak spot-size dependence suggests a quasi-ballistic phonon transport effect. Although the measured value demonstrates excellent heat conduction capability, it remains lower than the theoretical value expected for such a highly purified diamond, indicating phonon-scattering mechanisms beyond isotope scattering may remain relevant in this extreme isotope-purity regime. Considering the very low isotope and impurity levels in the present sample, vacancy scattering is a plausible source of phonon scattering and may limit the thermal conductivity of diamond. These findings suggest that, once isotope scattering is strongly suppressed, further enhancement of thermal conductivity requires great attention to vacancies.

**Conflict of Interest**

The authors declare that they have no conflict of interest.

**Acknowledgements**

J.L. and Z.C. thank Prof. Bo Sun for allowing them use his TDTR for spot verification. This work was supported by the National Natural Science Foundation of China (62574007, T2550270), and the National Key Research and Development Program of China (2024YFA1207901).

**Data availability**

The data that support the findings of this study are available from the corresponding authors upon reasonable request.